# Photonic Dirac Waveguides

Svetlana Kiriushechkina[1*], Anton Vakulenko[1*], Daria Smirnova[2*], Sriram Guddala[1], Filipp Komissarenko[1], Monica Allen[3], Jeffery Allen[3], Alexander B. Khanikaev[1]

[1]Electrical Engineering and Physics, The City College of New York (USA), New York, NY 10031, USA
[2]Research School of Physics, The Australian National University, Canberra ACT 2601, Australia
[3]Air Force Research Laboratory, Munitions Directorate, Eglin AFB, USA
* These authors contributed equally to this work

**Abstract**

The Dirac equation is a paradigmatic model that describes a range of intriguing properties of relativistic spin-½ particles, from the existence of antiparticles to Klein tunneling. However, the Dirac-like equations have found application far beyond its original scope, and has been used to comprehend the properties of graphene and topological phases of matter. In the field of photonics, the opportunity to emulate Dirac physics has also enabled topological photonic insulators. In this paper, we demonstrate that judiciously engineered synthetic potentials in photonic Dirac systems can offer physical properties beyond both the elementary/quasi-particles' and topological realms. Specifically, we introduce a new class of optical "Dirac waveguides", whose guided electromagnetic modes are endowed with pseudo-spin degree of freedom. Pseudo-spin coupled with the ability to engineer synthetic gauge potentials acting on it, enables control over the guided modes which is unattainable in conventional optical waveguides. In particular, we use a silicon nanophotonic metasurface that supports pseudo-spin degree of freedom as a testing platform to predict and experimentally confirm a spin-full nature of the Dirac waveguides. We also demonstrate that, for suitable trapping potentials, the guided modes exhibit spin-dependent field distributions, which gives rise to their distinct transport and radiative properties. Thereby, the Dirac waveguides manifest spin-dependent radiative lifetimes – the non-Hermitian spin-Hall effect – and open new avenues for spin-multiplexing, controlling characteristics of guided optical modes, and tuning light-matter interactions with photonic pseudo-spins.

**Introduction**

Recent progress in subwavelength nanopatterning of optical materials has enabled designer photonic systems that can emulate diverse and exciting physical phenomena[1,2]. Photonic structures with Dirac-like conical dispersion represent one class of designers' materials, that have led to the prediction and observation of unconventional modes in photonic graphene[3], weak antilocalization[4,5], topological phenomena[6–13], novel non-Hermitian[14–20] and nonlinear[21–27] physics.

The emulation of the Dirac physics is based on the introduction of additional lattice symmetries into the design of photonic structures. Typically, hexagonal symmetry is used because it enables diabolic Dirac points near high-symmetry points of the Brillouin zone in momentum space. In two dimensions, this allows valley physics in photonic modes exhibiting spinless Dirac-like dispersion[3,28,29]. Here the diabolic Dirac point plays the role of a magnetic monopole in momentum space and gives rise to a range of topological phenomena[30–32,10]. To emulate a true spin-full Dirac equation, the pseudo-spin degree of freedom can be added by either introduction of additional symmetries[33–35] or, alternately, via folding the valleys onto one another by lattice perturbations[36–39]. In this work, we utilize the second approach and demonstrate that the spin-full nature of Dirac photonic structures can be leveraged to create a new type of photonic systems that can trap and guide optical modes with an additional degree of control enabled by photonic pseudo-spin. Synthetic gauge potentials produced by nonuniform nano-patterning allow generation of spin-dependent field distributions and even spin-dependent non-Hermitian effects.

**Theoretical model**

To explore new Dirac physics, we consider a system described by a spin-full massive Dirac Hamiltonian of the form:

$$\widehat{H}(r) = \widehat{\sigma}_z \widehat{s}_z m(r) - i\widehat{\sigma}_x \widehat{s}_z \partial_x - i\widehat{\sigma}_y \widehat{s}_0 \partial_y \qquad (1)$$

where $\hat{s}_i, \hat{\sigma}_i$ are Pauli matrices for the pseudo-spin and sublattice degrees of freedom, respectively, and $m(r)$ is the position-dependent mass term. The Dirac waveguide is created by a non-uniform mass term $m(x)$ and the trapping "potential" in the $x$-direction is formed when a "guiding" region of smaller mass (compared to the "cladding" region $|m_g(x)| < |m_c|$) is present. Thus, a simple Dirac waveguide can be constructed by choosing a square-well-shaped mass term, where the confinement mechanism due to the wider bandgap for the cladding region is the most evident. Assuming the eigenstates of the Hamiltonian (1) in the form of guided modes, $|\psi(r)\rangle = u(x) \exp(i\tilde{k}_y y)$, where $\tilde{k}_y$ is the wavenumber in the direction of propagation of guided modes ($y$-direction), one can explicitly derive an expression[31] for new (squared in the $\hat{\sigma}_i$ subspace) Hamiltonian $\widehat{\mathcal{H}} = \widehat{H}^2$ of the form

$$\widehat{\mathcal{H}}(\tilde{k}_y) = -\partial_x^2 + [\tilde{k}_y^2 + m^2(x)] - i\widehat{\sigma}_y \widehat{s}_z [\partial_x m(x)]. \qquad (2)$$

The second term in this more familiar form (Helmholtz-like) of Hamiltonian Eq. (2) represents a well-recognizable 1D "potential well" for the guided modes in the waveguide and does not depend on the pseudo-spin of the mode. However, the third term clearly introduces such dependence showing the pseudo-spin dependent form of the eigenmodes of the Hamiltonians (2) and (1).

This spin-degenerate nature of the Dirac waveguides and the ability to generate spin-dependent potentials with carefully designed inhomogeneous distributions of the mass term, opens new opportunities to control light. For example, one can envision "spintronics for light" with spin-polarized photons in Dirac waveguides analogous to similar physics demonstrated by electrons in solids. One difference from electronic systems stems from the very nature of light and its ability to couple and leak into the radiative continuum with the related non-Hermitian physics. The spin-dependent potentials lead to distinct modal profiles for two photonic pseudo-spins, giving rise to different radiative properties and lifetimes of modes carrying opposite pseudo-spins. This latter effect is referred to as non-Hermitian spin-Hall effect.

We first analyze solutions of the Dirac equation (1) with the non-uniform, but symmetric distribution of the mass term shown in subplot Fig. 1a to illustrate the new physics enabled by the spin-full nature of the system and the non-Hermitian spin-Hall effect in particular. The spatially symmetric profile allows us to minimize the spin-dependence. The mechanism of mode confinement in this potential is clear and stems from the narrower photonic bandgap in the center of the structure compared to a broader bandgap on the sides (i.e., in the cladding). The Dirac eigenvalue problem for the quadratic potential well is solved numerically by applying the Susskind discretization scheme[40,41]. Our calculations confirm that the band structure contains additional bands separated from the gapped continuum when compared to the case of uniform mass term. These bands are doubly degenerate with respect to the pseudo-spin, obey particle-hole symmetry and thus appear at both positive and negative energies. The field profiles of these bands confirm that the modes are indeed localized at the center of the system with perfectly degenerate spectra for the two pseudo-spins. Anticipating the non-Hermitian nature of the experimental metasurface system, we endow the lower and upper bands with different radiative characteristics by assigning the lower (negative energy) and upper (positive energy) bands with quadrupolar and dipolar profiles at $\tilde{k}_y = 0$, respectively, by analogy to the leaky modes in nanostructured topological metasurfaces[36,39,42]. This allows us to calculate the complex-valued spectrum of the modes of the Dirac waveguide. The quality factor $Q = \text{Re}(\omega)/(2\text{Im}(\omega))$, which quantifies the radiative losses in the open system, is estimated as $Q = \omega_0 W/P$, where $\omega_0 \equiv Re(\omega)$ is the real part of the complex eigenfrequency, $W$ is the stored energy of a mode, and $P$ is the radiated power[43]. The radiated power in our phenomenological model is computed by Fourier transforming the 2D surface electric current distribution, associated with the dipolar component of the spinor, and integrating the out-of-plane Poynting flux for partial plane waves over the $\tilde{k}_x$ spectrum. In this approach, the electromagnetic fields are evaluated using the boundary conditions with the surface current density. The calculation confirms that in the case of a symmetric potential well, the imaginary part of the frequency, depicted by the quality factor (the color of the bands in Fig. 1), is identical for the two pseudo-spins.

Next, we consider an asymmetric distribution of the mass term. The results for the analytically solvable kink profile $m(x) = \tanh(x/w)$, corresponding to the Pöschl-Teller-like trapping potential[31,44] in the Schrödinger-like equation with Hamiltonian (2), are shown in Fig. 1b. The spectrum of bound states is given by $\omega_n = \pm\sqrt{\tilde{k}_y^2 + 1 - (1 - n/w)^2}$, where $n = 0,1,2 \ldots [w]$. Similar to the symmetric case, we observe emergence of two additional solutions that correspond to the two pseudo-spins at both positive and negative energies. In the asymmetric case, however, the field profiles show completely different field distributions, confirming that the two pseudo-spins perceive different effective trapping potentials. Insets to the band structure plot in Fig. 1b show profiles for two pseudo-spins with two specific values of momenta, $\tilde{k}_y > 0$ and its time-reversal partner $-\tilde{k}_y$, for positive and negative energies. The profiles are noticeably different for the two different pseudo-spins, with the maximum in the center for spin-up state $|\psi^\uparrow\rangle$ instead of minimum for spin-down $|\psi^\downarrow\rangle$ state for $\tilde{k}_y > 0$.

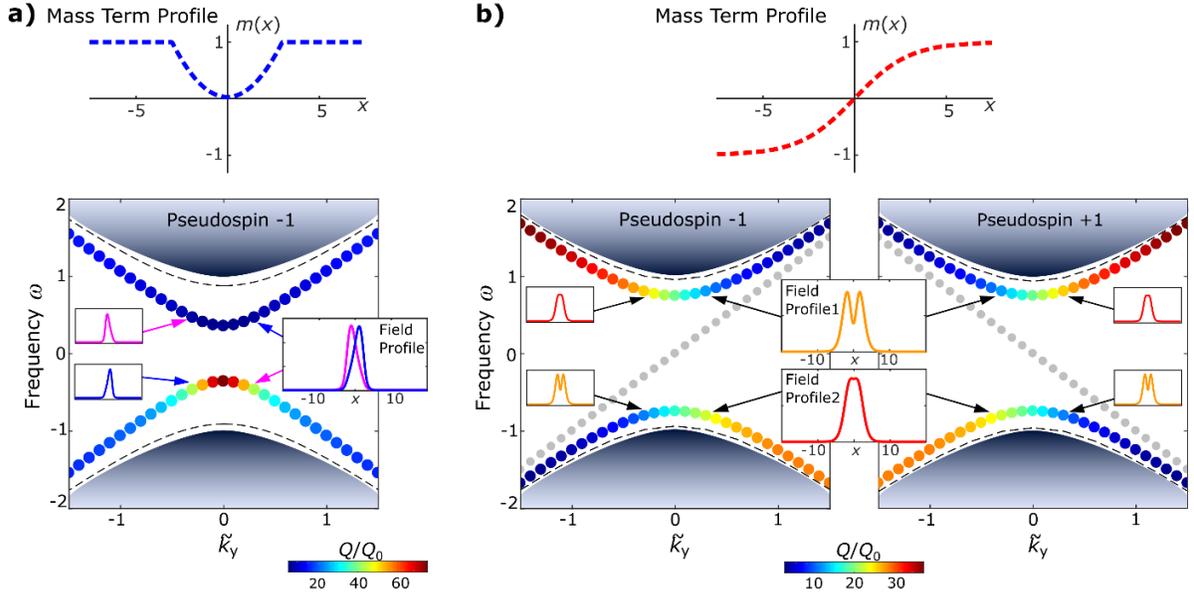

**Fig. 1. Spin-full guiding Dirac potentials generated by spatially variable mass term. a**, Symmetric mass term with parabolic shape in the center of the Dirac waveguide $m(x) = (x/w)^2$ of the width $w = 3$ (top panel) and corresponding spectrum of four doubly degenerate lowest energy spin-full guided modes (dotted line), four similar higher-energy states (dashed lines), and bulk continuum modes (bottom panel). The spectra of two pseudo-spins are doubly degenerate in both real and imaginary part, therefore only one spin is shown. The radiative quality factor of the modes is color-coded and breaks the particle-hole symmetry of the spectrum. The field profiles for different energy/frequency and momenta are shown as insets. **b**, Asymmetric mass term $m(x) = \tanh(x/w)$ corresponding to the Pöschl-Teller trapping potential (top panel) and corresponding energy/frequency spectra of the spin-polarized guided modes trapped near the center ($x = 0$). Two spins have different radiative lifetimes (non-Hermitian spin-Hall effect) color-coded by the quality factor $Q$ of the modes, and distinct field profiles (insets) due to spin-dependent character of the waveguide.

As expected from the time-reversal and particle-hole symmetries of this system, reversal of the momentum or reversal of the sign of energy leads to flipping of the field profiles for the two pseudo-spins. We note that in the case of asymmetric trapping profile, the mass term inversion leads to the formation of a topological domain wall that supports topological edge states with linear dispersion, $\omega_{\downarrow,\uparrow} = \pm \tilde{k}_y$. These modes are not of interest for this study and are plotted in gray in the figures below.

Another important observation that can be made from our calculations shown in Fig. 1b, is that, despite different field distributions, the modes guided by the Dirac waveguide retain their spin-degeneracy in the real part of the spectrum in the vicinity of $\tilde{k}_y = 0$. The radiative nature of the modes, however, implies that different field profiles will have dramatic effect on the imaginary part of the spectra. The calculations of the quality factors of the modes, shown in color, confirm the spin-dependent radiative properties and corroborate the non-Hermitian spin-Hall effect.

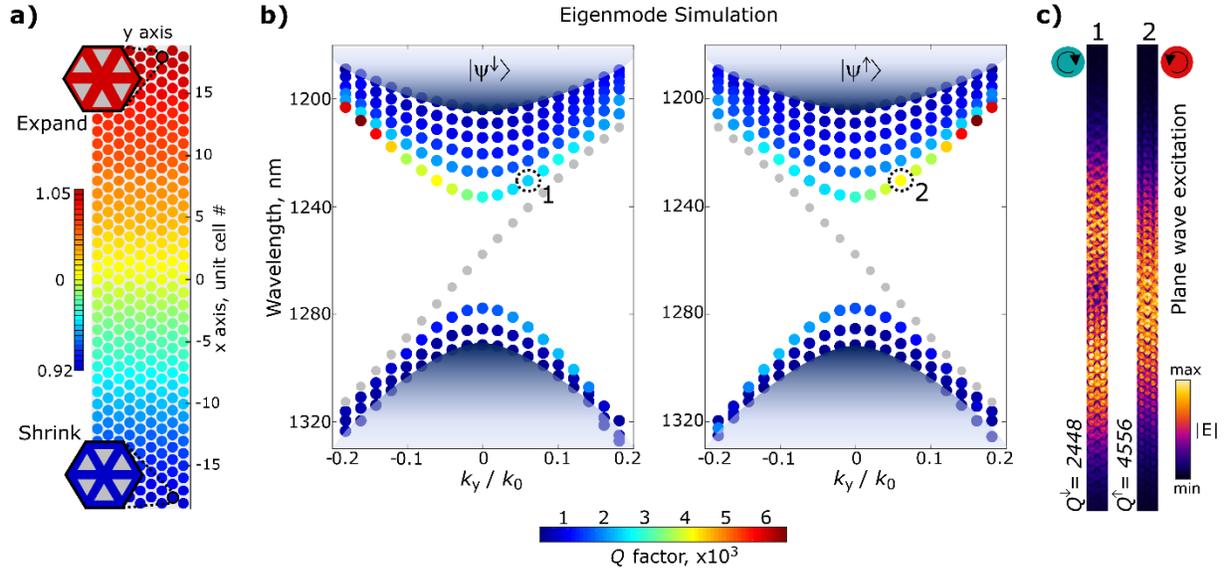

**Fig. 2. First-principles design of a spin-full Dirac meta-waveguide and non-Hermitian spin-Hall effect. a**, Dirac metasurface with variable lattice perturbation gives rise to the position-dependent effective mass term. The geometry of the resultant Dirac meta-waveguide changes from shrunken at the bottom to expanded at the top. The color bar shows the degree of perturbation from an ideal graphene-like lattice[37]. **b**, Complex eigenfrequency spectra of pseudo-spin-up and -down modes with lifetimes color-coded as radiative quality factors. **c**, Field distributions for Dirac waveguide modes excited by circularly polarized light of opposite helicities calculated for the same wavelength and momentum corresponding to the positions in the eigenfrequency spectra denoted by circles 1 and 2. Spin-dependent field distribution and the resultant difference in the radiative lifetimes (quality factors in **b**) of the modes highlight the non-Hermitian spin-Hall effect.

**First-principles studies**

We chose a leaky silicon photonic crystal (metasurface) design[38,37] with a pseudo-spin-degenerate Dirac cone at normal incidence ($k = 0$) to experimentally verify the predicted mechanism of light trapping and pseudo-spin-dependent properties of modes. The structure represents a lattice with unit cells containing a hexamer of triangles. This lattice structure exhibits a gap that opens when the symmetry of the structure is reduced by shifting triangles either towards (lower inset in Fig. 2a) or away from each other (upper insert in Fig. 1a), giving rise to trivial and topological phases, respectively. This design has been previously used to demonstrate spin-Hall-like topological photonic[38,45] and polaritonic phases[46–48]. Here, we are specifically interested in a new class of spin-full guided modes that can be induced by variation of the mass term in the spin-full Dirac Hamiltonian emulated by this system.

We create the Dirac waveguide by adiabatically varying the distance between the triangles that form the unit cell across one of the directions, as shown in Fig. 2a. This produces a linear mass-term profile that varies from negative to positive for the pseudo-spin-up $|\psi^\uparrow\rangle$ states, and positive to negative for the pseudo-spin-down $|\psi^\downarrow\rangle$ states of the 2D Dirac Hamiltonian. Our first-principle numerical simulations Fig. 2b confirm the formation of a new class of spin-degenerate guided eigenmodes, which split from the bulk continuum. The field profiles of these states show that they represent modes trapped by the effective potential produced by the mass-term variation. The profiles for pseudo-spin-up $|\psi^\uparrow\rangle$ and pseudo-spin-down $|\psi^\downarrow\rangle$ states of the same frequency and momentum, that correspond to the lowest energy states, are plotted in Fig. 2c. These profiles clearly highlight differences in the field distribution and confirm the spin-dependent character of the effective potential trapping these modes. More importantly, our first-principle calculations prove the non-Hermitian spin-Hall effect because the distinct field distributions give rise to different radiative characteristics and disparate radiative lifetimes and quality factors in modes of opposite pseudo-spins.

**Experimental results**

The guiding and the non-Hermitian spin-Hall effect in a spin-full Dirac meta-waveguide was experimentally confirmed using a structure with a variable mass term with the lattice perturbation as in Fig. 2a. The structure was fabricated on a silicon-on-insulator (SOI) substrate with a 220nm-thick silicon device layer patterned using electron beam lithography (EBL) followed by anisotropic reactive plasma etching (see Methods for details). The spectrum and propagation of the modes of the Dirac waveguide were probed with a custom-built microscope system that can image a sample in real and Fourier domains in the reflection geometry. To reveal the spin-dependent non-Hermitian physics of the structure, we utilized the circularly polarized nature of the far-field component of the modes supported by the system and performed direct selective excitation of the pseudo-spin-polarized guided modes using circularly polarized light focused on the center of the Dirac waveguide.

First, real-space images of the modes excited by the left- and right-circularly polarized light (of same intensity and momentum) were obtained, as shown in Fig. 3b top and bottom panels, respectively. The in-plane wavenumber $k_y$ was chosen by spatially filtering the incident beam with an aperture (see Methods for details). Figure 3b shows a clear difference in the excitation and propagation of the two guided modes of opposite pseudo-spins verifying the spin-dependent lifetimes of the pseudo-spin-up and pseudo-spin-down modes. Fitting of exponential decay of the guided modes yields a ratio of radiative lifetimes of pseudo-spin-down and pseudo-spin-up modes as 0.677 (see Supplementary Information, Section 1 for more details).

Next, we performed Fourier-plane (momentum-space) imaging of the structure for two cases of circularly polarized excitation. This allowed us to reconstruct a pseudo-spin-dependent band structure of the system. Top and bottom panels in Fig. 3c reveal a clear contrast in the visibility of the reflectivity spectra of the guided modes of two opposite pseudo-spins due to their different quality factors $Q^\uparrow(k_y) \neq Q^\downarrow(k_y)$ at any given value of wavenumber $k_y \neq 0$, as well as asymmetry with respect to the wavenumber inversion $Q^\uparrow(k_y) \neq Q^\uparrow(-k_y)$. We note that a relatively wide spread of the guided Dirac modes allows us to optimize the excitation efficiency by overlapping the spot size of the beam with the mode profile. This makes the guided Dirac modes especially pronounced compared to other states supported by the structure, such as more tightly localized topological edge state and delocalized bulk states, which are also seen in the experimental spectra.

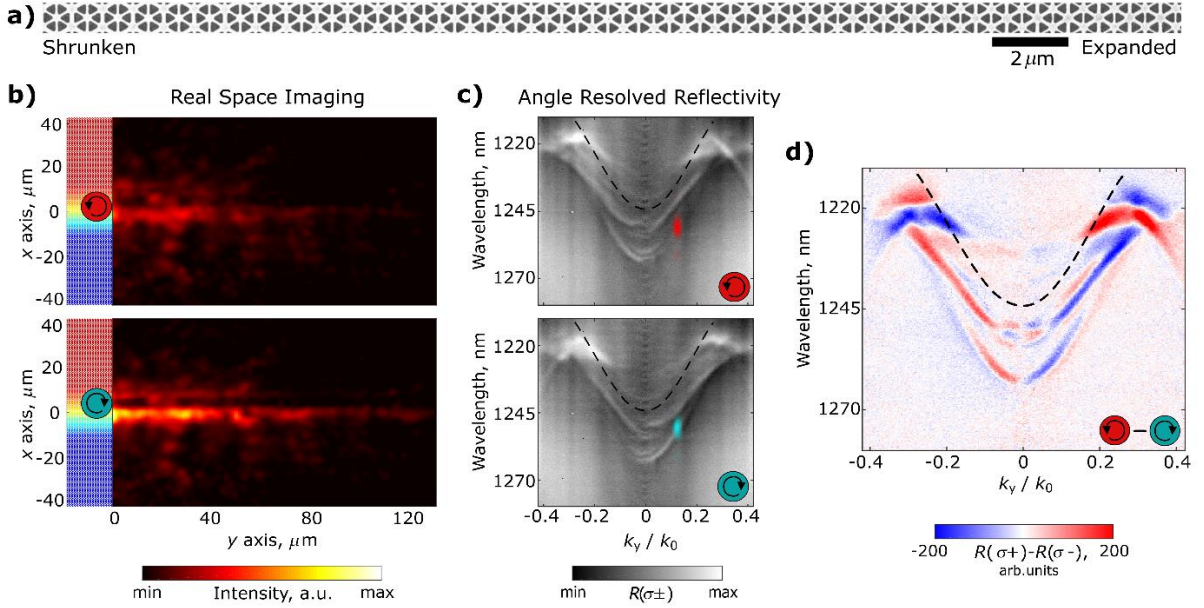

**Fig. 3. Experimental results evidencing pseudo-spin polarized transport and non-Hermitian spin-Hall effect in the Dirac meta-waveguide. a**, SEM image of a cross-section of the Dirac meta-waveguide with asymmetric mass term distribution and zero mass at the center. **b**, Real-space images of the spectrally and momentum-space filtered excitations of the lowest upper-frequency (positive energy) guided modes of the Dirac meta-waveguide, driven by left (top panel) and right (bottom panel) circularly polarized light for selective excitation of pseudo-spin-up and pseudo-spin-down modes, respectively. Non-Hermitian spin-Hall effect manifests in differing excitation efficiencies and propagation lengths for two guided modes of opposite pseudo-spins. **c**, Momentum-space (Fourier-plane) images of the Dirac meta-waveguide in the reflection geometry showing dispersion of the upper (positive energy) guided and bulk pseudo-spin-up and pseudo-spin-down modes. The colored dots indicate positions of spectral and momentum filtering for the real-space image in **a**. **d**, Differential reflectivity spectra evidencing contrast in excitation and leakage of the guided modes of opposite pseudo-spins. For comparison with results for the Dirac meta-waveguide with symmetric mass term distribution see Supplementary Fig. S2.

We note that no spin-dependence was observed for the symmetric waveguide (Supplementary Information, Section 2). The contrast between the cases of the two pseudo-spins is enhanced in Fig. 3d with a differential reflectivity spectrum $\Delta R(k_y) = R^\uparrow(k_y) - R^\downarrow(k_y)$, which provides additional confirmation of different radiative properties of the two spin-polarized modes and non-Hermitian spin-Hall effect.

**Conclusion**

In summary, here we demonstrate a new type of spin-full photonic waveguide based on Dirac metasurfaces. The underlying Dirac physics allows emulation of the spin degree of freedom – an on-chip photonic pseudo-spin – and generation of spin-dependent effective trapping potentials. As a particular example, we show through theoretical analysis and experimental measurements that a variation of the effective mass term in the Dirac metasurface allows trapping and guiding of pseudo-spin-degenerate guided modes to produce a spin-full Dirac meta-waveguide. We predict that inhomogeneous mass-terms give rise to distinct field distributions of the modes with opposite pseudo-spins. In the case of the leaky metasurface, the distinct field distributions yield different radiative lifetimes of pseudo-spin-polarized modes – a non-Hermitian spin-Hall effect of light. This effect is probed directly with spin-selective Fourier-plane imaging of the Dirac meta-waveguide.

The demonstrated pseudo-spin-dependent trapping and guiding of light opens a completely new direction for applications of photonic pseudo-spins, beyond topological concepts. We anticipate that the spin-full nature of the modes will find its use in on-chip quantum photonic devices where quantum information could be encoded by a photonic pseudo-spin. The spin-dependent field distributions and quality factors of the modes can also be used for selective (e.g., polarization dependent) control of light-matter interactions on a photonic chip, which envisions novel design principles of polaritonic, active, and nonlinear photonic devices.

**Methods**

**Numerical simulations.** We used commercial software COMSOL Multiphysics for the first-principles simulations presented in this work. The nanophotonic metasurface design was implemented in a silicon slab of 220 nm thickness on a glass substrate to yield a center of the bandgap at $\lambda$~1250 nm wavelength. The size of the unit cell with a hexamer of triangle-shaped air holes was $a_0$=680 nm, and the side of the equilateral triangle was $s$=260 nm. The metasurface with an adiabatically changing mass term was simulated using an array of 39 unit cells along the direction of the mass term profile variation aligned with $x$ axis. The degree of perturbation of the unit cell was linearly changed from the most expanded unit cell with $R = 1.05 \times a_0/3$, where $R$ is the distance between the center of unit cell to the centroid of each triangular hole at one edge of the array, to the unperturbed unit cell ($R = a_0/3$) in the center, and to the most shrunken unit cell ($R = 0.92 \times a_0/3$) at the opposite end of the array. The periodic boundary conditions were applied in the $y$ axis direction. All other boundaries were surrounded by perfectly matched layers. In the eigenfrequency spectra, modes were separated by the sign of pseudo-spin, which was calculated as the out-of-plane component of the angular momentum $L_z$ using the expression from work[49]. In the eigenmode simulations, the quality factor $Q$ of each mode was calculated from the complex-valued eigenfrequency $\omega_c$ as $Q = Re(\omega_c)/2Im(\omega_c)$. To confirm that the pseudo-spin polarized modes can be selectively excited by circularly polarized light, we performed frequency domain simulations with the incident plane wave source. The quality factor was then estimated from the resonant spectral response as $Q = \omega_0/\Delta\omega$, where $\omega_0$ is the central resonant frequency, and $\Delta\omega$ is a spectral width of the peak at half maximum.

**Sample fabrication.** The designed Dirac meta-waveguide was fabricated on the SOI substrates (220 nm of Si, 2 $\mu$m of buried oxide layer) with Electron Beam Lithography (Elionix ELS-G100). First, the substrates were spin-coated with e-beam resist ZEP520A-7 of approximately 150 nm thickness and then baked for 4 minutes at 180°C. For efficient charge dissipation a layer of anti-charging agent (DisCharge H20x2) of 50 nm thick was spin-coated on top of the resist. After e-beam lithography exposure anti-charging agent was removed by rinsing with DI water and the resist was developed in n-Amyl Acetate at 0°C for approximately 30 sec. Next, the silicon layer of the exposed area was vertically etched to the depth of 220 nm by inductively coupled plasma in the Oxford PlasmaPro System. A recipe of etching is based on C4F8/SF6 gases and reaches etching rate of about 2.5 nm/sec at 5°C table temperature. Finally, the residue of the resist was removed by the sample immersion into NMP solution heated to 60 °C.

**Experimental setup.** A custom-built near-infrared microscope was developed to image spectral dispersion of the Dirac metasurfaces. Light from a halogen lamp was collimated and focused to the sample surface using a long working distance 50X microscope objective (BoliOptics 50x, 0.42 NA). The back focal plane of the objective was imaged in 4$f$ configuration using the combination of a tube lens and a Fourier lens on to the entrance slit of the spectrometer (SpectraPro-HRS500, Teledyne Princeton Instruments). The dispersion from 300gr/mm grating was imaged on to the NIR camera (NIT HiPe SenS 640) connected at the exit slit of the spectrometer. A pair of linear polarizer and quarter wave plate were used in the incident optical path for the circular polarization studies. For the real space imaging of the Dirac waveguide modes, we used a laser beam with a linewidth of 5 nm. The laser beam was generated by supercontinuum light-source Leukos Electro-VIS with connected Leukos Tango-NIR2 acousto optic tunable filter. A telescope was built in the incident beam optical path with an aperture (25 $\mu$m) in the $k$-space position of the first lens. The selective wavevector ($k$) excitation was performed by translating the aperture along $k_y$-direction of the $k$-space to couple a specific waveguide mode. Image of real space propagation of the excited Dirac waveguide mode was captured by the NIR camera.

**Data availability**
The data that support the findings of this study are available from the corresponding author upon reasonable request.

**Acknowledgments**
The work was supported by the Office of Naval Research (ONR) award N00014-21-1-2092, the National Science Foundation (NSF) grant DMR-1809915, and the Simons Collaboration on Extreme Wave Phenomena. D. S. acknowledges support from the Australian Research Council (DE190100430). J. A. and M. A. would like to thank AFRL/RW Emerging Technologies for their support.

**Contributions**

S.K., A.V. and D.S. contributed equally to this work. A.K. conceived the research. D.S. performed theoretical calculations. S.K. and A.V. performed first-principle simulations, fabrication of samples and optical characterization, including real space imaging and angle-resolved reflectivity measurements. A.V., S.G. and F.K. assembled the experimental setup. A.K., M.A. and J.A. guided and supervised the project. All authors contributed to discussion of the results and manuscript preparation.

**Competing interests**

The authors declare no competing interests.

**References**

1. Qiao, W. *et al.* Toward Scalable Flexible Nanomanufacturing for Photonic Structures and Devices. *Adv. Mater.* **28**, 10353–10380 (2016).

2. Li, C. *et al.* Dielectric metasurfaces: From wavefront shaping to quantum platforms. *Prog. Surf. Sci.* **95**, 100584 (2020).

3. Plotnik, Y. *et al.* Observation of unconventional edge states in 'photonic graphene'. *Nat. Mater.* **13**, 57–62 (2014).

4. Bliokh, K. Y. Weak antilocalization of ultrarelativistic fermions. *Phys. Lett. Sect. A Gen. At. Solid State Phys.* **344**, 127–130 (2005).

5. Morozov, S. V. *et al.* Strong suppression of weak localization in graphene. *Phys. Rev. Lett.* **97**, (2006).

6. Lu, L., Joannopoulos, J. D. & Soljačić, M. Topological photonics. *Nat. Photonics* **8**, 821–829 (2014).

7. Wang, Z., Chong, Y., Joannopoulos, J. D. & Soljačić, M. Observation of unidirectional backscattering-immune topological electromagnetic states. *Nature* **461**, 772–775 (2009).

8. Hafezi, M., Demler, E. A., Lukin, M. D. & Taylor, J. M. Robust optical delay lines with topological protection. *Nat. Phys.* **7**, 907–912 (2011).

9. Cheng, X. *et al.* Robust reconfigurable electromagnetic pathways within a photonic topological insulator. *Nat. Mater.* **15**, 542–548 (2016).

10. Khanikaev, A. B. & Shvets, G. Two-dimensional topological photonics. *Nat. Photonics* **11**, 763–773 (2017).

11. Ozawa, T. *et al.* Topological photonics. *Rev. Mod. Phys.* **91**, 015006 (2019).

12. Slobozhanyuk, A. *et al.* Three-dimensional all-dielectric photonic topological insulator. *Nat. Photonics* **11**, 130–136 (2017).

13. Yang, Y. *et al.* Realization of a three-dimensional photonic topological

insulator. *Nature* **565**, 622–626 (2019).

14. Szameit, A., Rechtsman, M. C., Bahat-Treidel, O. & Segev, M. PT-symmetry in honeycomb photonic lattices. *Phys. Rev. A - At. Mol. Opt. Phys.* **84**, 021806R (2011).

15. Zhen, B. *et al.* Spawning rings of exceptional points out of Dirac cones. *Nature* **525**, 354–358 (2015).

16. Leykam, D., Bliokh, K. Y., Huang, C., Chong, Y. D. & Nori, F. Edge Modes, Degeneracies, and Topological Numbers in Non-Hermitian Systems. *Phys. Rev. Lett.* **118**, 076801 (2017).

17. Liu, T. *et al.* Second-Order Topological Phases in Non-Hermitian Systems. *Phys. Rev. Lett.* **122**, 076801 (2019).

18. Liu, Y. G. N., Jung, P. S., Parto, M., Christodoulides, D. N. & Khajavikhan, M. Gain-induced topological response via tailored long-range interactions. *Nat. Phys.* **17**, 704–709 (2021).

19. Xia, S. *et al.* Nonlinear tuning of PT symmetry and non-Hermitian topological states. *Science* **372**, 72–76 (2021).

20. Parto, M., Liu, Y. G. N., Bahari, B., Khajavikhan, M. & Christodoulides, D. N. Non-Hermitian and topological photonics: Optics at an exceptional point. *Nanophotonics* **10**, 403–423 (2020).

21. Lumer, Y., Plotnik, Y., Rechtsman, M. C. & Segev, M. Self-localized states in photonic topological insulators. *Phys. Rev. Lett.* **111**, 243905 (2013).

22. Hadad, Y., Khanikaev, A. B. & Alù, A. Self-induced topological transitions and edge states supported by nonlinear staggered potentials. *Phys. Rev. B* **93**, 155112 (2016).

23. Leykam, D. & Chong, Y. D. Edge Solitons in Nonlinear-Photonic Topological Insulators. *Phys. Rev. Lett.* **117**, 143901 (2016).

24. Hadad, Y., Soric, J. C., Khanikaev, A. B. & Alú, A. Self-induced topological protection in nonlinear circuit arrays. *Nat. Electron.* **1**, 178–182 (2018).

25. D'Aguanno, G. *et al.* Nonlinear topological transitions over a metasurface. *Phys. Rev. B* **100**, 214310 (2019).

26. Smirnova, D., Leykam, D., Chong, Y. & Kivshar, Y. Nonlinear topological photonics. *Appl. Phys. Rev.* **7**, 021306 (2020).

27. Maczewsky, L. J. *et al.* Nonlinearity-induced photonic topological insulator.

*Science* **370**, 701–704 (2020).

28. Gao, F. *et al.* Topologically protected refraction of robust kink states in valley photonic crystals. *Nat. Phys.* **14**, 140–144 (2018).

29. Liu, J. W. *et al.* Valley photonic crystals. *Adv. Phys.: X* **6**, (2021).

30. Haldane, F. D. M. & Raghu, S. Possible realization of directional optical waveguides in photonic crystals with broken time-reversal symmetry. *Phys. Rev. Lett.* **100**, 013904 (2008).

31. Raghu, S. & Haldane, F. D. M. Analogs of quantum-Hall-effect edge states in photonic crystals. *Phys. Rev. A* **78**, 033834 (2008).

32. Rechtsman, M. C. *et al.* Photonic Floquet topological insulators. *Nature* **496**, 196–200 (2013).

33. Khanikaev, A. B. *et al.* Photonic topological insulators. *Nat. Mater.* **12**, 233–239 (2013).

34. Silveirinha, M. G. Bulk-edge correspondence for topological photonic continua. *Phys. Rev. B* **94**, (2016).

35. Bisharat, D. J. & Sievenpiper, D. F. Electromagnetic-Dual Metasurfaces for Topological States along a 1D Interface. *Laser Photonics Rev.* **13**, 1900126 (2019).

36. Wu, L. H. & Hu, X. Scheme for achieving a topological photonic crystal by using dielectric material. *Phys. Rev. Lett.* **114**, 223901 (2015).

37. Barik, S., Miyake, H., Degottardi, W., Waks, E. & Hafezi, M. Two-dimensionally confined topological edge states in photonic crystals. *New J. Phys.* **18**, 113013 (2016).

38. Barik, S. *et al.* A topological quantum optics interface. *Science* **359**, 666–668 (2018).

39. Gorlach, M. A. *et al.* Far-field probing of leaky topological states in all-dielectric metasurfaces. *Nat. Commun.* **9**, 909 (2018).

40. Tworzydło, J., Groth, C. W. & Beenakker, C. W. J. Finite difference method for transport properties of massless Dirac fermions. *Phys. Rev. B - Condens. Matter Mater. Phys.* **78**, 235438 (2008).

41. Hernández, A. R. & Lewenkopf, C. H. Finite-difference method for transport of two-dimensional massless Dirac fermions in a ribbon geometry. *Phys. Rev. B* **86**, 155439 (2012).

42. Smirnova, D. *et al.* Third-Harmonic Generation in Photonic Topological Metasurfaces. *Phys. Rev. Lett.* **123**, 103901 (2019).

43. Vučković, J., Lončar, M., Mabuchi, H. & Scherer, A. Optimization of the Q factor in photonic crystal microcavities. *IEEE J. Quantum Electron.* **38**, 850–856 (2002).

44. L. D. Landau and E. M. Lifschitz, Quantum Mechanics: Non-Relativistic Theory (Pergamon 1977).

45. Parappurath, N., Alpeggiani, F., Kuipers, L. & Verhagen, E. Direct observation of topological edge states in silicon photonic crystals: Spin, dispersion, and chiral routing. *Sci. Adv.* **6**, (2020).

46. Liu, W. *et al.* Generation of helical topological exciton-polaritons. *Science* **370**, 600–604 (2020).

47. Li, M. *et al.* Experimental observation of topological Z2 exciton-polaritons in transition metal dichalcogenide monolayers. *Nat. Commun.* **12**, 4425 (2021).

48. Guddala, S. *et al.* Topological phonon-polariton funneling in midinfrared metasurfaces. *Science.* **374**, 225–227 (2021).

49. Bliokh, K. Y., Bekshaev, A. Y. & Nori, F. Optical momentum and angular momentum in complex media: From the Abraham-Minkowski debate to unusual properties of surface plasmon-polaritons. *New J. Phys.* **19**, 123014 (2017).

# Supplementary Information: Photonic Dirac Waveguides

Svetlana Kiriushechkina[1*], Anton Vakulenko[1*], Daria Smirnova[2*], Sriram Guddala[1], Filipp Komissarenko[1], Monica Allen[3], Jeffery Allen[3], Alexander B. Khanikaev[1]

[1]Electrical Engineering and Physics, The City College of New York (USA), New York, NY 10031, USA
[2]Research School of Physics, The Australian National University, Canberra ACT 2601, Australia
[3]Air Force Research Laboratory, Munitions Directorate, Eglin AFB, USA
* These authors contributed equally to this work

**Section 1. The quality factor of guided modes from experiment**

The quantitative comparison of spin-dependent lifetimes of the guided modes in the Dirac meta-waveguide can be evaluated from the images of experimental real-space propagation of the guided modes. Fitting of an exponential decay of the measured reflectance along the direction of propagation is shown in the Fig.S1 for two excitations of opposite handedness and the same intensity. Fitting function for intensity $I$ is given as:

$$I = R^2 \exp(-2\alpha y), \tag{1}$$

where $R^2$ is the amplitude (or the brightness), $\alpha$ is the attenuation constant (the imaginary component of the wavenumber of the mode), and $y$ is the position along the waveguide. Since the propagation of the leaky guided modes in the Dirac meta-waveguide is associated with two main mechanisms of losses (i) the spin-dependent radiative loss $P_{rad}^{\uparrow,\downarrow}$ due to the radiative character of the modes and (ii) the scattering loss $P_{sc}$ originating from the defects and imperfections in the fabricated structure (e.g., due to the Rayleigh scattering), the quality factor $Q^{\uparrow,\downarrow}$ of the mode can be written as:

$$1/Q^{\uparrow,\downarrow} = 1/Q_{rad}^{\uparrow,\downarrow} + 1/Q_{sc}. \tag{2}$$

On the other hand, $Q$-factor can be evaluated as:

$$Q = \omega\tau/2, \tag{3}$$

where $\omega$ is the frequency of the mode and $\tau$ is its lifetime. The latter can be related to the group velocity $v_{gr}$ and the propagation length $L$ as:

$$\tau = L/v_{gr} = 1/\alpha v_{gr}. \tag{4}$$

Substituting Eq. (4) into Eq.(3), it yields the relation between $Q$ and $\alpha$:

$$Q = \omega/\alpha v_{gr}. \tag{5}$$

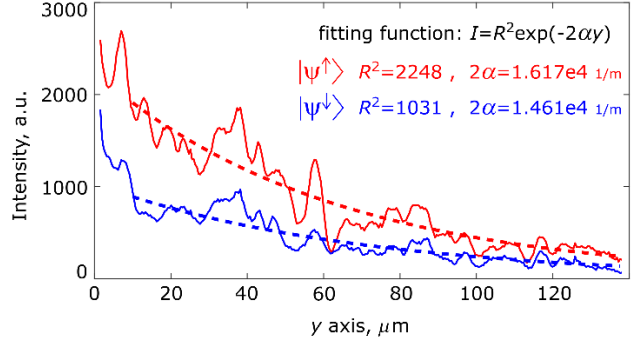

**Figure S1. Fitting the real space propagation data for the guided modes of opposite pseudo-spins**. Fitting to an exponential decay (dashed line) yields the brightness $R^2$ and attenuation $\alpha$ parameters of the guided Dirac modes. Data for the pseudo-spin-up mode is shown in red color and for the pseudo-spin-up mode - in blue color.

The group velocity is defined as $v_{gr} = \frac{\partial \omega}{\partial k}$, and for the case of *k*-selective excitation of the Dirac guided mode ($k_y/k_0 = 0.13$) it was estimated from Fourier space image as $v_{gr} \approx \frac{\Delta\omega}{\Delta k} = 0.129c$.

Parameters $\alpha^\downarrow$ and $\alpha^\uparrow$ were extracted from fitting and give propagation lengths $L^\downarrow = (124 \pm 7)\ \mu m$ $L^\uparrow = (137 \pm 9)\ \mu m$ and quality factors $Q^\downarrow = 2408 \pm 129$ and $Q^\uparrow = 2663 \pm 167$. The values of $Q^\downarrow$ and $Q^\uparrow$ are very close, which implies (from Eq. 2) that the *Q*-factor of the mode is largely defined by spin-independent scattering ($Q_{sc} \ll Q_{rad}^{\uparrow,\downarrow}$), which represents the dominant loss mechanism.

It can be clearly seen from real-space images of the mode propagation that the pseudo-spin-down mode is much darker than the pseudo-spin-up mode. Considering that a power of excitation beam and a background noise level were the same for both measurements the brightness coefficient $R_{\uparrow,\downarrow}^2$ in Eq.(1) can be written approximately as:

$$R_{\uparrow,\downarrow}^2 \sim \left(\frac{\frac{1}{Q_{rad}^{\uparrow,\downarrow}}}{\frac{1}{Q_{rad}^{\uparrow,\downarrow}}+\frac{1}{Q_{sc}}}\right)^2 \cong \left(\frac{Q_{sc}}{Q_{rad}^{\uparrow,\downarrow}}\right)^2. \qquad (6)$$

Considering that $Q_{sc}$ is the same for both pseudo-spin-up and pseudo-spin-down modes, the ratio of brightness parameters equals to the inverse ratio of radiative *Q*-factors:

$$\frac{R_\uparrow}{R_\downarrow} \cong \frac{Q_{rad}^\downarrow}{Q_{rad}^\uparrow}. \qquad (7)$$

From extracted fitting parameters we obtain $Q_{rad}^\downarrow/Q_{rad}^\uparrow = 0.677 \pm 0.020$ which is consistent with the results of our numerical simulations $\tilde{Q}^\downarrow/\tilde{Q}^\uparrow = 0.537$.

**Section 2. Experimental results for the meta-waveguide with symmetric trapping potential**

To confirm the absence of the spin-dependence in the case of the symmetric trapping potential, we fabricated the respective sample with the variation of the perturbation from the graphene lattice (expanded case) shown in Fig. S2a. The reflectivity appears to be very symmetric with respect to the spin and momentum (Fig. S2b), yielding a reflectivity difference (Fig. S2d) negligible compared to that of the asymmetric mass term distribution of the main text (Fig. 3d).

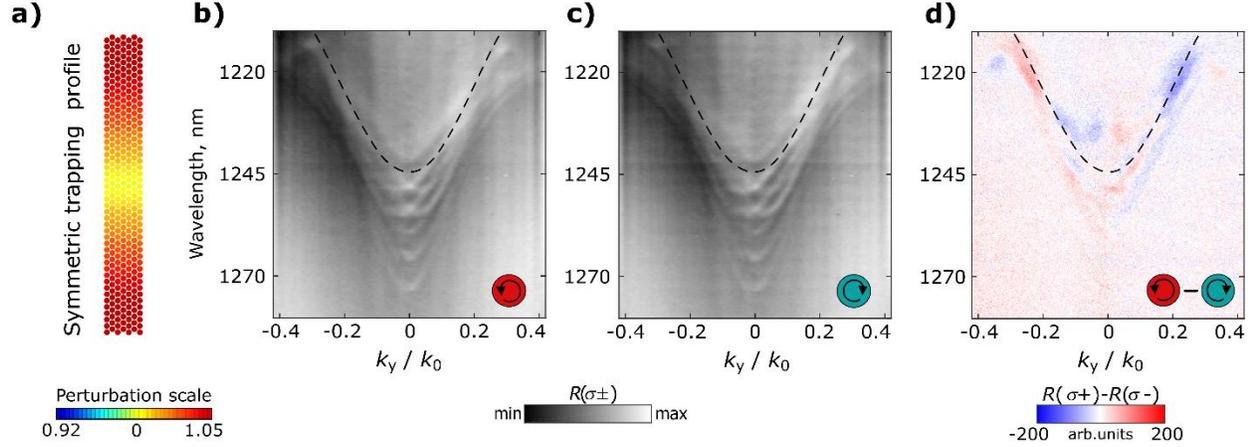

**Figure S2. Angle-resolved reflectivity maps for the Dirac meta-waveguide with symmetric trapping potential**. **a**, Schematic geometry of the symmetric meta-waveguide. **b-c**, Angle-resolved reflectance measurements of the Dirac meta-waveguide showing dispersion of the upper (positive energy) guided and bulk pseudo-spin-up and pseudo-spin-down modes for the circular left and right polarization, respectively. **d**, Differential reflectivity spectra which shows negligible contrast in the excitation and leakage of the guided modes of opposite pseudo-spins for the case of symmetric meta-waveguide. The limits of the colorbar are kept the same as in Fig. 3d of the main text for comparison.